\documentclass[conference]{IEEEtran}
\IEEEoverridecommandlockouts

\usepackage{cite}
\usepackage{amsmath,amssymb,amsfonts}
\usepackage{algorithmic}
\usepackage{graphicx}
\usepackage{textcomp}
\usepackage{xcolor}
\usepackage[acronym]{glossaries}
\usepackage{xspace}  
\usepackage[hidelinks]{hyperref}
\usepackage{booktabs}
\usepackage{xfrac}
\usepackage{subcaption}

\def\BibTeX{{\rm B\kern-.05em{\sc i\kern-.025em b}\kern-.08em
		T\kern-.1667em\lower.7ex\hbox{E}\kern-.125emX}}

\newcommand{\uppaal}{{\sc Uppaal}\xspace}

\newcommand{\stratego}[0]{\uppaal~{\sc Stratego}\xspace}

\newacronym{tg}{TG}{Timed Game}
\newacronym{pn}{PN}{Petri Net}
\newacronym{mpc}{MPC}{Model Predictive Control}
\newacronym{mld}{MLD}{Mixed Logical Dynamical}
\newacronym{sumo}{SUMO}{Simulation of Urban MObility}
\newacronym{traci}{TraCI}{Traffic Control Interface}
\newacronym{tl}{TL}{time loss}
\newacronym{fc}{FC}{fuel consumption}
\newacronym{milp}{MILP}{Mixed-Integer Linear Programming}

\begin{document}

\title{A Flow-Efficient and Legal-by-Construction Real-Time
 Traffic Signal Control Platform
}

\author{\IEEEauthorblockN{Frederik Baymler Mathiesen\IEEEauthorrefmark{2}}
\IEEEauthorblockA{\textit{Department of Computer Science} \\
\textit{Aalborg University}\\
Aalborg, Denmark \\
fmathi16@student.aau.dk}
\and
\IEEEauthorblockN{Garey Fleeman}
\IEEEauthorblockA{\textit{EECS Department} \\
\textit{University of California, Berkeley}\\
Berkeley, United States \\
garey\_fleeman@berkeley.edu}
}

\maketitle

\begin{abstract}
Inefficiencies in traffic flow through an intersection lead to stopping vehicles, unnecessary congestion, and increased accident risk. In this paper, we propose a traffic signal controller platform demonstrating the ability to increase traffic flow for arbitrary intersection topologies. This model uses Model Predictive Control on a Mixed Logical Dynamical system to control the state of independently controlled traffic signals in a single intersection, removing constraints forcing the selection of signals from a set of phases. Further, we use constraints to impose a guarantee on the output of the system to be in the set of permissible actions under constraints including precise yellow timing, minimum inter-lane green transition timing, and selection of signal states with non-conflicting dependencies. We evaluate our model on a simulated 4-way intersection and an intersection in Denmark with true traffic data and a currently implemented timing schedule as a baseline. Our model shows at least 22\% reduction in time loss compared to baseline light schedules, and timing shows that this system can feasibly run online predictions at a frequency faster than 2s/prediction optimizing over a prediction horizon of 25s.
\end{abstract}


\section{Introduction}
Traffic signals form an essential infrastructure for coordinating traffic in congested areas with the benefits of a small footprint, high throughput, and starvation management. 
However, these benefits come at a cost: interrupted traffic flow, which implies stopping vehicles, congestion, and increased accident risk\cite{french2006}. 
A common solution to increase throughput and reduce number of stops is to coordinate multiple intersection into a green wave or using adaptive signal control where the controller adapts to the current traffic situation.
In this paper, we present a platform for adaptive signal control for a single intersection using \gls{mpc} on an \gls{mld} system.
To evalutate the efficiency and effectiveness, we run experiments in \gls{sumo} on two different intersections: a 4-way intersection and a simulation of a real intersection in Denmark.

Kamal et al.\cite{kamal2012} propose modelling the traffic flow using hybrid dynamics of discrete and continuous variables, namely an \gls{mld} system, and controlling using \gls{mpc}.
They model macroscopic traffic with multiple intersections that affect each other by sending traffic through connected road segments.
The traffic volumes allow fractional vehicles, which is an artifact of an assumption that using large scale sensor networks to achieve complete accuracy about the vehicle dynamics was impossible.
Today, we believe that a combination of GPS tracking and traffic radars yield a close to complete image of the traffic, and we assume a perfect knowledge about the position and velocity of vehicles.

Eriksen et al.\cite{eriksen2017} solve the control problem using \glspl{tg} and reinforcement learning in \stratego where the road users constitutes the environment and the agent is the traffic signal controller.
Their solution includes online controller synthesis based on the current traffic flow and simulations for a finite prediction horizon.
One contribution is to optimize on a model of complete vehicle trajectories rather than conditions on point measurement such as induction loops.

\begingroup\renewcommand\thefootnote{\IEEEauthorrefmark{2}}
\footnotetext{Research conducted while visiting University of California, Berkeley}
\endgroup

The control problem is heavily constrained by law and traffic safety considerations such as green interval between conflicted movements, minimum green times to safely cross the intersection, and yellow period \cite{beeber}.
Additionally, the rules are not consistent across jurisdictions\footnote{In Europe, the yellow and red lights are lit simultaneously (called amber) as an intermediate state in the transition from red to green, while in the United States, the transition happens directly.}, which poses a problem for the generalizability of any traffic signal controller. 

We propose a legal-by-construction \gls{mpc} platform with an \gls{mld} model for traffic flow.
The controller is parametric to different intersections and jurisdictions, and allows the traffic light operator to weigh different optimization objectives like queue size, number of stops, and throughput.
We only optimize a single intersection but with hidden synchronization, i.e. traffic lights synchronize with the platoons of vehicles such that the benefit of two adjacent, traffic controlled intersections is higher than the sum of the individual benefits \cite{hansen2017}.

The controller computes a schedule online, and thus, the computation time is paramount for the controller. We assume each time step is 1s. Therefore, the controller must compute a new schedule in less than 1s. The requirement can be relaxed, e.g. by using a schedule for 2s, but for optimal performance, a new schedule should be computed regularly. 

Section \ref{sec:control} explains the plant dynamics and constraints along with a definition of \gls{mpc}. Section \ref{sec:experiments} explains the experiment methodology and measures, and Section \ref{sec:results} presents the empirical results. Finally, Section \ref{sec:discussion} discusses the results, implications, and future work.


\section{Control}\label{sec:control}
   

    We analyze the intersection by considering a set of $n$ independently controlled signals. A single traffic signal in the intersection can correspond to multiple lanes, e.g. two parallel lanes are going straight based on the same light. 
    In traditional traffic light control, there exists a set of legal light state combinations referred to as phases, and the phases are switched simultaneously \cite{chandler2013signalized}. By separating signals, we remove restrictions on the controller for choosing from these phases, allowing the model to consider each lane independently.
    
    We will now define system and environment evolution, as well as the formal mathematical constraints imposed on the system and the cost function used for optimization, guaranteeing the system output is drawn from the set of legal phases\footnote{Equations are represented using the Hadamard operator $\odot$, but in practice, these represent linear constraints for each element of the vectors.}.
    \subsection{System Definitions}
    
    We define a set of binary color states for all lanes as $\delta_g^{(k)}, \delta_y^{(k)}, \delta_r^{(k)}, \delta_a^{(k)}, \delta_{ng}^{(k)} \in \{0,1\}^n$. $\delta_{ng}^{(k)}$ represents not-green and is the sum of the red, yellow, and amber states.
    
    We calculate arriving vehicles, denoted $c_a^{(k)} \in \mathbb{N}^n$, and the maximum flow of lanes in a signal is denoted $f_m \in \mathbb{N}^n$. In our experiments, $c_a^{(k)}$ is simulated using SUMO, and $f_m$ is constant and found using simulation during green light states.
    
    To restrict solutions to those that do not have intersecting traffic flows, we define a constant conflict matrix $C \in \{0,1\}^{n \times n}$. Element $C_{ij}$ indicates that allowing traffic on both lane $i$ and $j$ is permissible. $C$ is symmetric and $C_{ii}=0$ for $1 \leq i \leq n$ since a lane cannot be in conflict with itself.
    
    We define yellow and amber time period vectors denoted $t_{p_a}, t_{p_y} \in \mathbb{N}^n$ respectively. These vectors represent the amount of time each signal must remain constant in the amber or yellow state before a transition.
    Similarly, we impose a minimum time between a green transition of two conflicting lanes, which may be asymmetric.
    Consider a left turn in lane $i$ conflicts with a straight path from lane $j$.
    Before transitioning from an active $\delta_{g[j]}^{(k)}$ to an active $\delta_{g[i]}^{(k+1)}$, we may need to only wait a small amount of time for the vehicles to exit the lane, but the opposite transition may take longer to allow cars waiting for a left turn to clear the intersection.
    We define a green interval matrix as $T_{m_g} \in \mathbb{N}^{n \times n}$ with element $T_{m_g[ij]}$ indicating the amount of time between the end of a green light in $i$, and the beginning of a green light in $j$.
    \subsection{Plant State Evolution}
    We will now describe the plant dynamics used to update the state variables.
    \begin{align}
        \intertext{\textbf{Flow: }We define the traffic flow to be $f^{(k)} \in \mathbb{N}^n$, which is the number of cars passing through the intersection in each lane. We compute the flow as follows:}
        &f^{(k)}=\min{\left(f_m, \delta_g^{(k)}\odot\left(c_a^{(k)}+q^{(k-1)}\right)\right)}
        \intertext{The minimum operation ensures the flow is not be larger than the sum of previous queue and current arrivals.}
        \intertext{\textbf{Queue: }We count the number of cars waiting in the queue, referred to as $q^{(k)} \in \mathbb{N}^n$. We update the queue using the previously calculated flow and current arrivals:}
        &q^{(k)} = q^{(k-1)} +c_a^{(k)} - f^{(k)}
        \intertext{\textbf{Timers: }We track timers for light states along with the maximum wait time, $t_g^{( k)}, t_y^{(k)}, t_a^{(k)}, t_{ng}^{(k)}, t_w^{(k)} \in \mathbb{N}^n$. To evolve the timer states, we must considering the current light state. For color timers $c\in\{g, y, a, ng\}$, we use:}
        &t_c^{(k)} = \left(t_c^{(k-1)}+\vec{1}\right) \odot \delta_c^{(k)}
        \intertext{The wait time vector also includes an indicator function on the queue to only increment for lanes with vehicles waiting.}
        &t_w^{(k)} =  \left(t_w^{(k-1)}+\vec{1}\right) \odot \delta_{ng}^{(k)} \odot I_{\geq 1}(q^{(k)})
        \intertext{\textbf{Stops: } We define the number of stops to be the number of cars arriving at a non-green light denoted as $s^{(k)} \in \mathbb{N}^n$. The stops are computed by comparing the arriving vehicles to the lights with a non-green state:}
        &s^{(k)} = \delta_{ng}^{(k)} \odot c_a^{(k)}
    \end{align}
    \subsection{System Constraints}
    We now discuss constraints guaranteeing permissible actions. Note that we use equality and comparison operators elementwise. We have excluded implementation-specific constraints in the system and only present those which verify guarantee of not making illegal actions.
    \begin{align}
        \intertext{\textbf{Single Light: }To constrain each lane to a single active light state, we impose $n$ linear constraints using:}
        &\delta_g^{(k)}+\delta_r^{(k)}+\delta_y^{(k)}+\delta_a^{(k)}=\vec{1}
        \intertext{\textbf{Conflict: }Applying the constraint matrix $C$ to the blocking lanes  $b^{(k)}=\delta_g^{(k)}+\delta_y^{(k)}+\delta_a^{(k)}$, we get a quadratic constraint\footnotemark:}
        &b^{(k)T}Cb^{(k)} = 0
        \intertext{\textbf{Permissible Transition: }To block illegal light transitions (such as red to yellow), we use $8n$ linear constraints where $\delta_{c_{p}}^{(k-1)}$ is the previous light state, and $\delta_{c_{b}}^{(k)}$ is the current state to be blocked:}
        &\delta_{c_{p}}^{(k-1)} + \delta_{c_{b}}^{(k)} \leq \vec{1}
        \intertext{\textbf{Yellow and Amber Timing: }Both the amber light state and yellow light state must be precisely timed. For $c \in \{y, a\}$, we $8n$ constraints of the form:}
        &t_{p_c} \odot \left(\delta_c^{(k-1)}-\delta_c^{(k)}\right) \leq t_{c}^{(k-1)}\label{time_const_1} \\
        &t_{p_c} \odot \delta_c^{(k)} \geq t_{c}^{(k)} \label{time_const_2}
        \intertext{Equation \ref{time_const_1} forces $\delta_c^{(k)}$ to remain the same as $\delta_c^{(k-1)}$ until the timer $t_c^{(k-1)} \geq t_{p_c}$, then become inactive. Equation \ref{time_const_2} become active when $t_c^{(k-1)} \geq t_{p_c}$ and forces a change to the next state.}
        \intertext{\textbf{Minimum Green Transition Interval: }To ensure the minimum green interval is met for any lane changing its green state to active, we compare the not-green times for all other lanes, and the entry in the green interval matrix for the lanes being compared. This is represented in matrix notations as:}
        &\left(T_{m_g}-t_{ng}^{(k-1)}\;\vec{1}^T\right) \odot \left(\vec{1}\;\delta_g^{(k)T}\right) \leq \vec{0}\;\vec{0}^T
    \end{align}
    This matrix equation represents $n^2$ quadratic constraints of which the diagonal $n$ constraints are trivially satisfied
    \footnotetext{For performance, this quadratic constraint can be split into linear constraints using the method in \cite{BEMPORAD1999407}.}.
    \subsection{Objective Function}
    We employ a linear objective function composed of a set of plant states $X^{(k)}=\{q^{(k)}, t_{w}^{(k)}, s^{(k)}, -f^{(k)}\}$ and controlled variables $U^{(k)}=\{\delta_{ng}^{(k)}\}$ as: 
    \begin{align}
       & O\left(X^{(k)}, U^{(k)}\right) = \sum_{d^{(k)} \in X^{(k)} \cup U^{(k)}}w_d \left(d^{(k)} \cdot \vec{1}\right)
    \end{align}
    where the terms $w_x$ represent the weight of cost term $x$. Note that the flow is represented with a negative sign because the objective is being minimized, but we intend to maximize flow.
    \subsection{Model Predictive Control}
\begin{figure}
    \centering
    \includegraphics[width=0.8\linewidth]{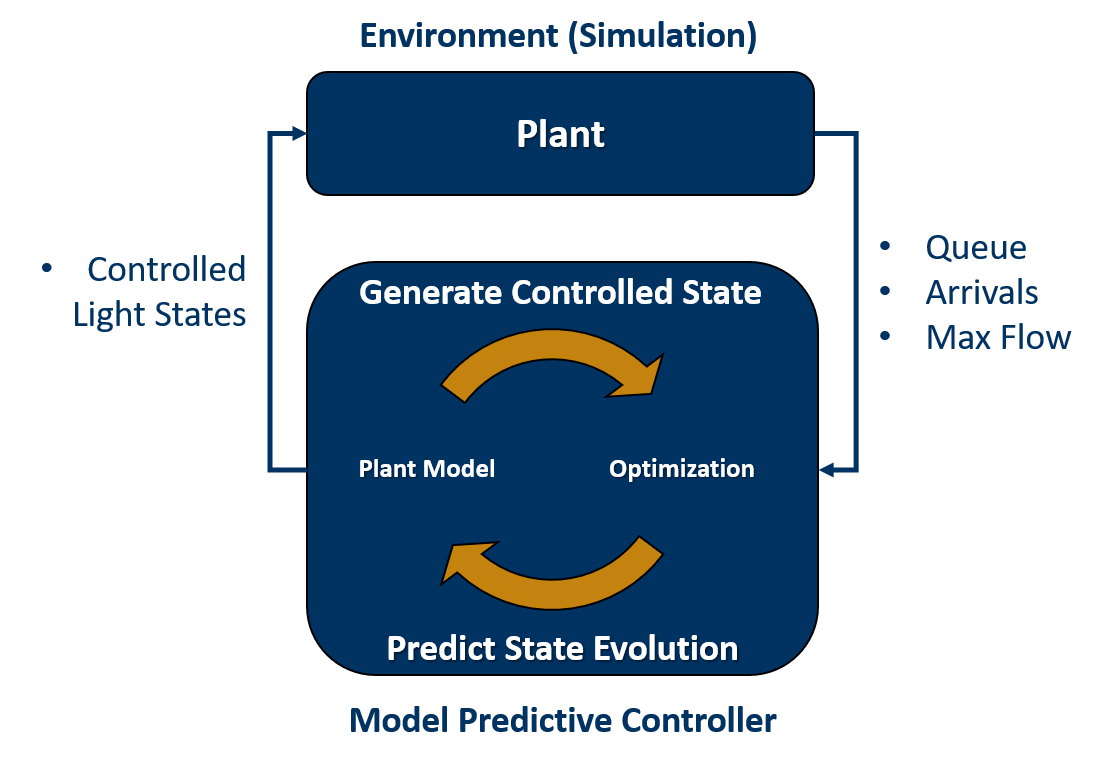}
    \caption{The MPC model and plant in a loop.}
    \label{fig:mpc_diagram}
\end{figure}

    \Gls{mpc} allows the system to consider the long-term cost of changing the traffic light state. 
    In \gls{mpc}, the controller optimizes the plant output over a future window, see Figure \ref{fig:mpc_diagram}. The controller solves the problem over the window by combining the constraints and objective function for each the time step into one \gls{milp} problem for the entire prediction horizon.
    To limit the control space but account for future implications, the controller can use a control horizon less than the prediction horizon.

\section{Experiments}\label{sec:experiments}
\begin{figure}
    \centering
    \includegraphics[width=0.85\linewidth]{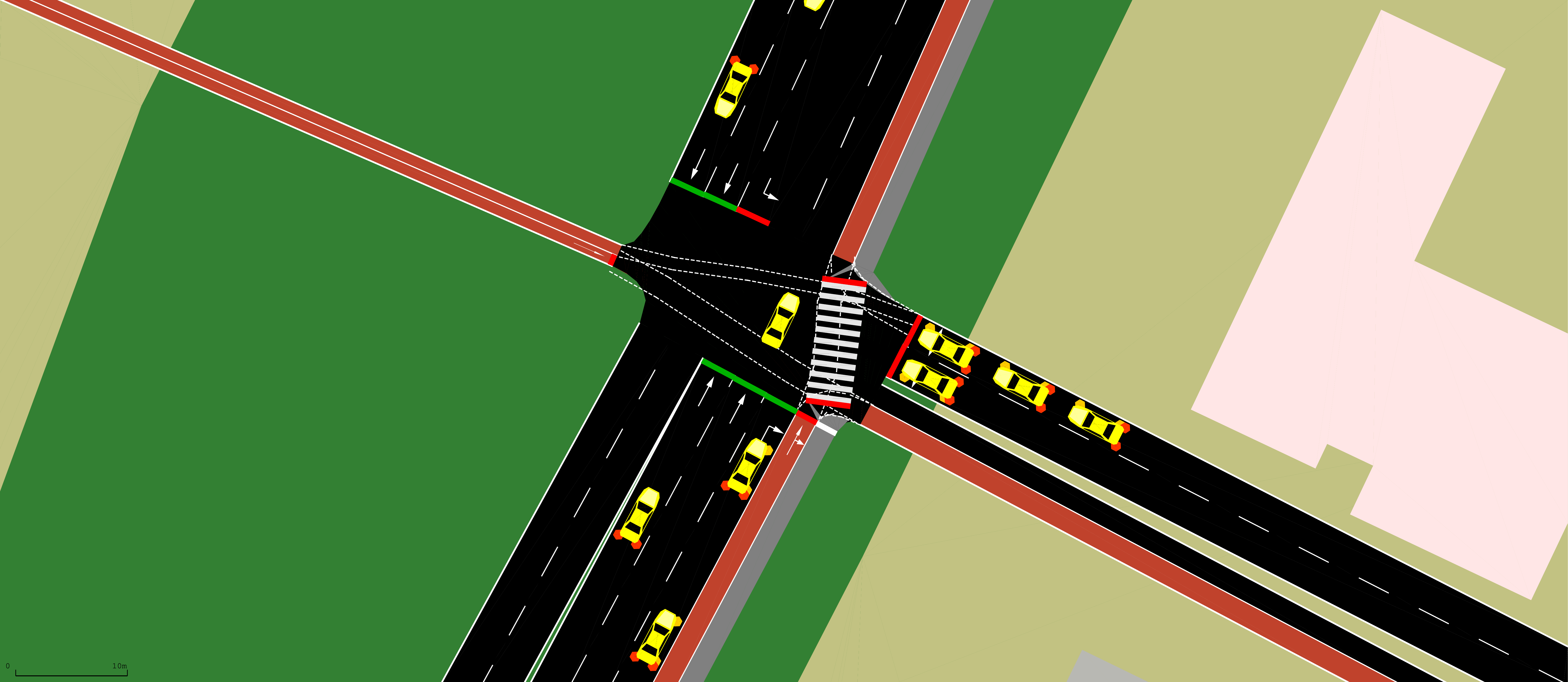}
    \caption{The Grenåvej/Egå Havvej intersection in SUMO. Red lanes are for bikes and grey is for pedestrians.}
    \label{fig:aarhus_maps}
\end{figure}

To evaluate the performance, we integrate our controller with \gls{sumo}, an open source, microscopic traffic simulator \cite{SUMO2018}, i.e. it simulates individual vehicle dynamic.
The experiments are conducted with binomially distributed traffic flow, which approximates a Poisson distribution \cite{Wenpin2019}, using \gls{sumo}'s demand generation. 
To test the system under various conditions, we construct 2 scenarios:
\begin{itemize}
    \item A 4-way intersection with symmetric traffic safety values, see Table \ref{tab:traffic_safety_parameters}. The probability of a vehicle arriving each second is $\sfrac{1}{6}$ for East/West and $\sfrac{1}{12}$ for North/South.
    \item The Grenåvej/Egå Havvej intersection in Denmark where we have access to the traffic safety parameters, the time program, and the true traffic flows throughout the day. Figure \ref{fig:aarhus_maps} is a map of the intersection.
\end{itemize}

\begin{table}
\caption{Parameters for the 4-way intersection.}
\begin{subtable}{0.45\linewidth}
    \caption{Traffic safety parameters.}
    \label{tab:traffic_safety_parameters}
    \centering
    \begin{tabular}{lr}
        \toprule
         Parameter &  \\
         \midrule
         Amber time & 0s \\
         Yellow time &  4s \\
         Minimum green time & 6s \\
         Green interval & 6s \\
         \bottomrule
    \end{tabular}
\end{subtable}
\hfill
\begin{subtable}{0.45\linewidth}
    \caption{Time program.}
    \label{tab:time_programs}
    \centering
    \begin{tabular}{lrr}
        \toprule
        Time & N/S & E/W \\
        \midrule
        20s & Green & Red \\
        4s & Yellow & Red \\
        2s & Red & Red \\
        40s & Red & Green \\
        4s & Red & Yellow \\
        2s & Red & Red \\
        \bottomrule
    \end{tabular}
\end{subtable}
\end{table}

For the 4-way intersection, we fabricate a time program as a baseline, see Table \ref{tab:time_programs}, and for the Aarhus intersection, we use the time program running in the intersection as a baseline.
    

For estimating the arrival time of each car, we assume a constant velocity, which is does not account for cars accelerating or decelerating, but it is sufficient since cars travel the majority at constant velocity, i.e. the speed limit. 

We evaluate the system with the average and 95\textsuperscript{th} percentile \gls{tl}, i.e. time lost by driving less than the ideal speed, and average \gls{fc}, which are measured by \gls{sumo}.
In addition to the effectiveness measures, we measure the execution time to compute a schedule during each of these experiments.

\section{Results}\label{sec:results}

\begin{table}
    \caption{Time loss and fuel consumption by controller and intersection; smaller is better. All \gls{mpc} controller experiments have a prediction horizon $P = 30$.}
    \label{tab:results_effectiveness}
    \centering
    \begin{tabular}{lrrr}
        \toprule
        Controller & Avg. TL [s] & 95\textsuperscript{th} perc. TL [s] & Avg. FC [ml] \\ %
        \midrule
        \multicolumn{4}{l}{\textbf{4-way intersection}} \\
        Time program & 20.58 & 48.34 & 76.97 \\
        MPC, $C = 15$ & 18.40 & 49.16 & 78.07 \\
        MPC, $C = 20$ & 16.09 & 35.32 & 76.18 \\
        MPC, $C = 25$ & 15.95 & 33.67 & 76.06 \\
        MPC, $C = 30$ & 15.96 & 33.81 & 76.07 \\
        \midrule
        \multicolumn{4}{l}{\textbf{Grenåvej/Egå Havvej intersection}} \\
        Time program & 23.30 & 62.20 & 110.07 \\
        MPC, $C = 15$ & 14.93 & 59.59 & 103.80 \\
        MPC, $C = 20$ & 13.53 & 44.30 & 102.54 \\
        MPC, $C = 25$ & 12.84 & 40.48 & 101.99 \\
        MPC, $C = 30$ & 12.96 & 43.48 & 102.06 \\
        \bottomrule
    \end{tabular}
\end{table}

\subsection{Time Loss \& Fuel Consumption}
From Table \ref{tab:results_effectiveness}, we see that a control horizon of 15 improve the avg. TL for both intersections but with a small penalty for the 95\textsuperscript{th} percentile TL. Every MPC controller with a control horizon of 20 or more is superior to the timed control, but with diminishing returns beyond $C = 20$, and worse effectiveness for $C = 30$ compared to $C = 25$.
Looking at individual lanes in Table \ref{tab:results_aarhus_by_lane}, we see a significant performance improvement for all lanes except pedestrians and bikes from West. 

For $C \geq 20$, fuel saved is approx. 1\% for the 4-way intersection and 7\% for the Grenåvej/Egå Havvej intersection.

\begin{table}
    \caption{\% improvement of TL and FC by signal of the Grenåvej/Egå Havvej intersection for the \gls{mpc} controller with $P = 30$ and $C = 20$ compared to the time program. Avg. FC is not listed for signals with only bicyclists and pedestrians.}
    \label{tab:results_aarhus_by_lane}
    \centering
    \begin{tabular}{lrrr}
        \toprule
        Signal & Avg. TL & 95\textsuperscript{th} perc. TL & Avg. FC \\
        \midrule
        North straight & 47.5\% & 38.5\% & 4.4\% \\
        North left turn & 32.3\% & 13.9\% & 13.2\% \\
        South straight & 49.5\% & 43.9\% & 5.9\% \\
        South right turn & 68.8\% & 55.2\% & 13.7\% \\
        South bike & 46.1\% & 48.2\% & - \\
        South pedestrian & -1.4\% & -7.0\% & - \\
        East & 27.5\% & 12.8\% & 7.7\% \\
        West & 6.9\% & -13.4\% & - \\
        \midrule
        Total & 40.0\% & 24.8\% & 8.6\% \\
        \bottomrule
    \end{tabular}
\end{table}

\subsection{Execution time}
For solving the \gls{milp} of the \gls{mpc}, we use Gurobi, a state of the art solver with support for mixed integer quadratic programming.
The experiments are conducted on a PC with an Intel\textregistered~Core\texttrademark~i7-4720HQ processor and 8 GB RAM.

\begin{figure}
    \centering
    \includegraphics[width=0.85\linewidth]{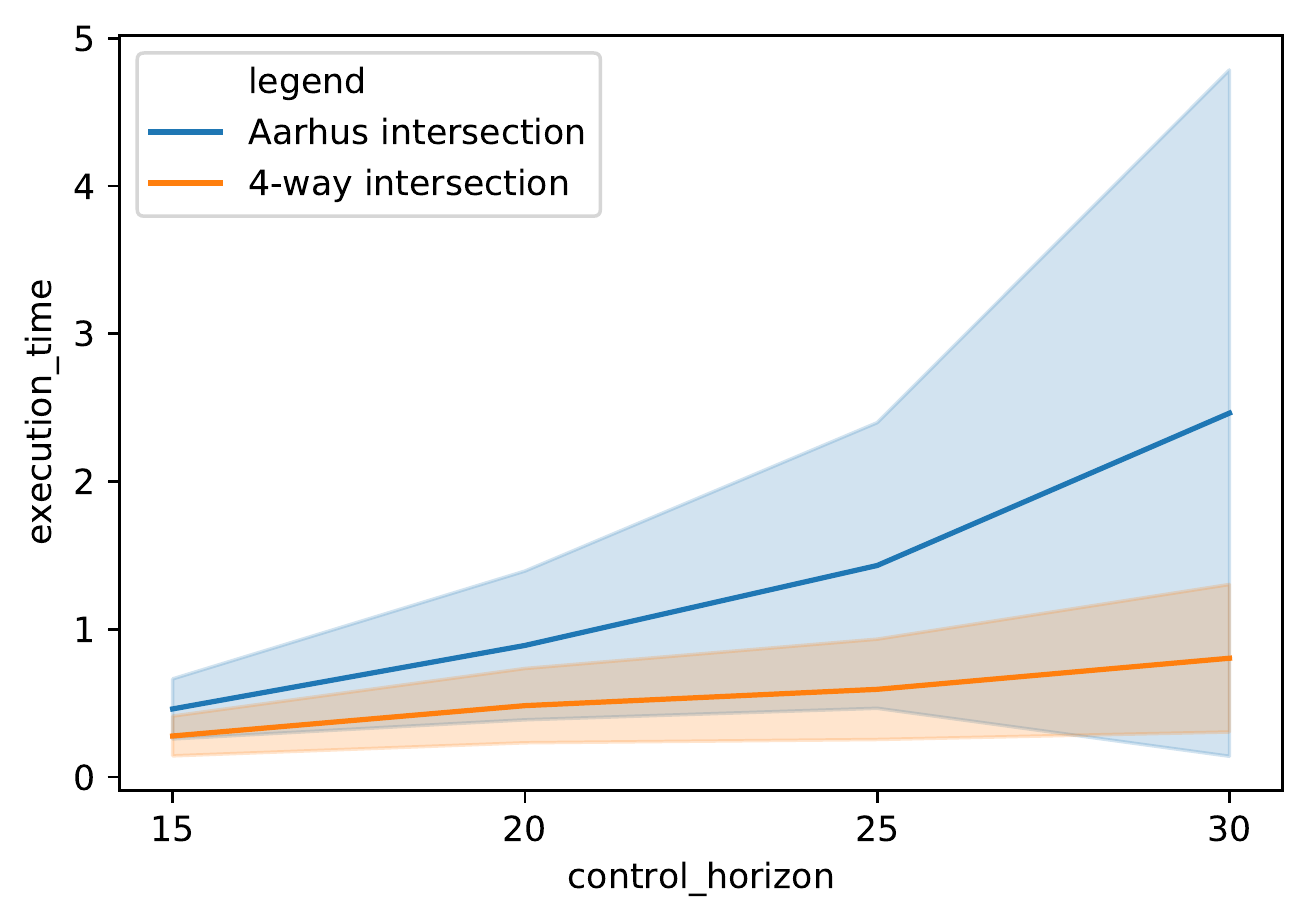}
    \caption{Execution time as a function of control horizon in second. The semi-transparent area is the standard deviation.}
    \label{fig:execution_time}
\end{figure}

As we can see on Figure \ref{fig:execution_time}, the execution time for the 4-way intersection is significantly less than 1s for a control horizon of both 15s and 20s time steps.
For a control horizon of 25s and 30s, one standard deviation is near 1s.
For the Grenåvej/Egå Havvej intersection, the mean for a control horizon of 20 being near 0.9s, which is too much for a production environment but within reasonable bounds for what is doable with runtime optimizations and better hardware.

\section{Discussion}\label{sec:discussion}
We have designed a traffic signal controller platform based on \gls{mpc}, which for feasible control horizons, i.e. approximately 1s computation time, reduces time loss more than 30\% for most lanes and fuel consumption more than 5\%.
The 95\textsuperscript{th} percentile is often reduced but at most 14\% higher, which can solved with per lane weights for the cost function.


To improve the effectiveness while maintaining feasibility, a better model of vehicle dynamics to predict arrival at the intersection would be a solution.
Future work can also focus on learning the arrival dynamics of individual intersection.

In conclusion, we have designed a legal-by-construction model predictive traffic signal controller showing a significant improvement compared to baseline timing schedules using real traffic data. This \acrlong{mpc} framework uses live traffic data to minimize an objective function, weighing several costs over a prediction horizon. The constraints on the controller guarantee only legal actions are taken and their design is agnostic to the traffic topology.

\bibliography{bibliography}{}

\begin{thebibliography}{1}

\bibitem{french2006}
L.~J. French and M.~S. French, ``Benefits of signal timing optimization and its
  to corridor operations,'' tech. rep., The Pennsylvania Department of
  Transportation, Smithfield, PA, 2006.

\bibitem{kamal2012}
M.~Kamal, J.~Imura, A.~Ohata, T.~Hayakawa, and K.~Aihara, ``Control of traffic
  signals in a model predictive control framework,'' {\em IFAC Proceedings
  Volumes}, vol.~45, no.~24, 2012.
\newblock 13th IFAC Symposium on Control in Transportation Systems.

\bibitem{eriksen2017}
A.~Eriksen, C.~Huang, J.~Kildebogaard, H.~Lahrmann, K.~Larsen, M.~Muniz, and
  J.~Taankvist, ``Uppaal stratego for intelligent traffic lights,'' in {\em
  12th ITS European Congress}, ERTICO - ITS Europe, 2017.

\bibitem{beeber}
J.~Beeber, ``Report to ctcdc on minimum yellow light change interval timing for
  signalized intersections,'' tech. rep., Safer Streets L.A.

\bibitem{hansen2017}
M.~Hansen, A.~Eriksen, J.~Taankvist, K.~Larsen, and H.~Lahrmann, ``Optimering
  af signalstyring i realtid: Intelligent styring af signalregulerede kryds ved
  anvendelse af maskinl{\ae}ring og objektdetektering,'' vol.~2017, Division
  for Transportation Engineering, AAU, 2017.

\bibitem{chandler2013signalized}
B.~E. Chandler, M.~Myers, J.~E. Atkinson, T.~Bryer, R.~Retting, J.~Smithline,
  J.~Trim, P.~Wojtkiewicz, G.~B. Thomas, S.~P. Venglar, {\em et~al.},
  ``Signalized intersections informational guide,'' tech. rep., United States.
  Federal Highway Administration. Office of Safety, 2013.

\bibitem{BEMPORAD1999407}
A.~Bemporad and M.~Morari, ``Control of systems integrating logic, dynamics,
  and constraints,'' {\em Automatica}, vol.~35, no.~3, 1999.

\bibitem{SUMO2018}
P.~A. Lopez, M.~Behrisch, L.~Bieker-Walz, J.~Erdmann, Y.-P. Fl{\"o}tter{\"o}d,
  R.~Hilbrich, L.~L{\"u}cken, J.~Rummel, P.~Wagner, and E.~Wie{\ss}ner,
  ``Microscopic traffic simulation using {SUMO},'' in {\em The 21st IEEE
  International Conference on Intelligent Transportation Systems}, IEEE, 2018.

\bibitem{Wenpin2019}
W.~Tang and F.~Tang, ``The poisson binomial distribution -- old \& new,'' 2019.

\end{thebibliography}
\bibliographystyle{ieeetr}

\end{document}